\pdfoutput=1

\documentclass[a4paper,12pt]{iopart}
\usepackage{graphicx}
\usepackage{color}
\usepackage{amsfonts}
\usepackage{amssymb}
\usepackage{wasysym}
\usepackage{cite}
\usepackage[super]{nth}
\usepackage{srcltx}
\begin{document}
\title{Revising the Musical Equal Temperament}

\author{Haye Hinrichsen}
\address{Universit\"at W\"urzburg, Fakult\"at f\"ur Physik und Astronomie \\Campus S\"ud, Am Hubland, 97074 W\"urzburg, Germany}

\ead{\tt hinrichsen@physik.uni-wuerzburg.de}

\begin{abstract}
Western music is predominantly based on the equal temperament with a constant semitone frequency ratio of $2^{1/12}$. Although this temperament has been in use since the \nth{19} century and in spite of its high degree of symmetry, various musicians have repeatedly expressed their discomfort with the harmonicity of certain intervals. Recently it was suggested that this problem can be overcome by introducing a modified temperament with a constant but slightly increased frequency ratio. In this paper we confirm this conjecture quantitatively. Using entropy as a  measure for harmonicity, we show numerically that the harmonic optimum is in fact obtained for frequency ratios slightly larger than $2^{1/12}$. This suggests that the equal temperament should be replaced by a harmonized stretched temperament as a new standard.
\end{abstract}

\vspace{2pc}

\maketitle

\pagestyle{plain}
\def\d{{\rm d}}
\renewcommand{\texttrademark}{${^{\mbox{\tiny\textregistered}}}$}
\section{Introduction}

\begin{figure}
\centering\includegraphics[width=155mm]{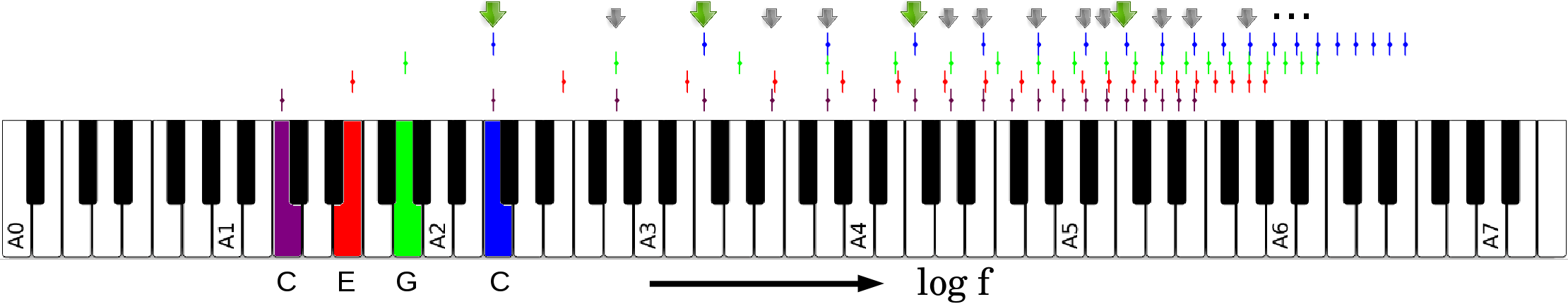}
\caption{\scriptsize Harmonicity of a trichord. Playing a C-Major trichord (C-E-G-C) one generates four series of partials which are shown above the keyboard at the corresponding positions. Note that neighboring partials have a constant \textit{frequency difference} while neighboring keys on the keyboard are characterized by a constant \textit{frequency ratio}. This explains why the distance between the partials decreases as we go to the right, demonstrating the mathematical incommensurability of the exponential scale and the linear partial series. The trichord is perceived as harmonic because various partials match, as indicated by the arrows. The large green arrows stand for perfect matching due to pure octaves. In the standard equal temperament, on which this figure is based, there are also various approximate coincidences, as marked by the smaller gray arrows.
}
\label{loglin}
\end{figure}

Musical intervals between two tones are perceived as harmonic if the corresponding frequencies are related by simple fractional ratios~\cite{Prout}. For example, the perfect fifth and the perfect fourth corresponds to the frequency ratios $\frac{3}{2}$ and $\frac{4}{3}$. Our ability to hear and identify such fractional frequency ratios is related to the fact that the natural spectrum of musical sounds consists not only of the fundamental frequency~$f_1$ but also involves a large number higher \textit{partials} (overtones) at integer multiple frequencies $f_n=n f_1$. Our sense of hearing is capable to detect matching partials, signaling us the impression of consonance and harmony. In other words, perceiving a perfect fifth as consonant does not mean that our ears like the numerical value $\frac{3}{2}$, it rather reflects the circumstance that the third partial of the lower tone coincides with the second of the upper (see Fig.~\ref{loglin}). 

A musical scale is a periodic system of notes ordered by increasing fundamental frequency. The Western \textit{chromatic scale}, on which we will focus in the present work, consists of twelve semitones per octave. As a musical scale is constructed in repeating patterns of intervals, it is always exponentially organized in the frequencies. For example, in the chromatic scale the frequency doubles from octave to octave. Aiming for a harmonic perception, musical scales are tuned in such a way that the fundamental frequencies capture or at least approximate simple fractional frequency ratios. However, in setting up such a tuning scheme one is immediately confronted with the problem that the exponential structure of the scale and the linear organization of the partials are incommensurable. 

The probably most natural example of such a tuning system is the \textit{just intonation}, where all intervals are built on simple fractional ratios with respect to a certain reference key. Music played in this reference key sounds very harmonic if not even sterile, but when played in a different key the same piece can be terribly out of tune. With the increasing complexity of music, however, frequent key changes became more important. Historically this led to the fascinating development of so-called \textit{temperaments}~\cite{TuningSystems}, i.e. tuning systems seeking for a reasonable compromise between harmonicity and invariance under transposition, attempting to reconcile the exponential structure of the scale with the linear organization of the partials. This development culminated in the so-called \textit{equal temperament} (ET) with a constant semitone ratio of $2^{1/12}$. This temperament is perfectly invariant under key changes, but except for the octave all intervals are out of perfect tune. The ET is the standard temperament of Western music and has been in use since the beginning of the \nth{19} century.

Despite the high degree of symmetry, some musicians continue to express their concern about certain intervals in the ET which exhibit unpleasant beats. This discontent may have contributed to the revival of historical temperaments, accompanied by a subculture of newly invented unequal temperaments, but with all these approaches the main achievement of the ET, namely its beautiful key invariance, is lost.

Is it possible to improve the chromatic ET without destroying its symmetry under key changes? As we will see in the present work, this is indeed possible. To understand the basic idea let us recall that the standard ET is determined by three conditions, namely, (i) its exponential structure of constant semitone intervals, (ii) twelve semitones per octave, and (iii) the obvious condition that the frequency should double on each octave. This means that the octave, the simplest and most fundamental of all intervals, is the only pure one in the ET, producing a static sound without any beats. But why? As we are ready to tolerate beats for any other interval, why not for octaves? By allowing the octave to exhibit beats, would it be possible to arrive at a more balanced temperament? 

In the past few decades there have been several proposals to study \textit{stretched equal temperaments}. In these temperaments the semitone frequency ratio is still constant but slightly larger than $2^{1/12}$. To my knowledge the first example of this kind was the Stopper\texttrademark \ tuning scheme introduced by B. Stopper in 1988, which favors a pure duodecime consisting of 19 semitones instead of the octave~\cite{Stopper}. In 1995 a similar suggestion based on a perfect fifth was put forward by S.~Cordier and K. Gillessen~\cite{Cordier,Gillessen,Gillessen2}, who argued that in practice such temperaments would be already in use. More recently A. Capurso suggested what is known as the \textit{circular harmonic system} (c.ha.s\texttrademark)~\cite{Capurso,Chiriano}, where all intervals exhibit beats. All these proposed chromatic temperaments belong to the same family in so far as they are perfectly equal, differing only in their semitone frequency ratio. However, while Stopper and Cordier simply replace the constraint of one pure interval by another, the c.ha.s temperament is based on a more complex reasoning and seeks for an optimal balance in a regime where none of the intervals is pure. 

Which of these temperaments establishes the best harmonic compromise? In this paper we would like to address this question quantitatively. To this end we apply the concept of \textit{entropy} as a measure for the harmonicity of a temperament. In statistical physics and information theory, entropy is known to quantify the degree of disorder in a data set. In Ref.~\cite{EPTPaper} we observed that the entropy of pure (harmonic) intervals is locally minimal since the entropy decreases when higher partials overlap. This suggests that the entropy of overlaid power spectra can be used as a measure of harmonicity. Very recently we were able to demonstrate that this concept can even be used for piano tuning~\cite{EPT}. Therefore, the obvious question would be what entropy can tell us about different stretched equal temperaments. 

In order to address this question, we generate artificial harmonic power spectra for various temperaments and compute the corresponding entropy numerically. As we will see below, our results confirm that the standard equal temperament has to be replaced by a stretched one and also suggests a range for the possible optimum.

It is very important to realize that stretched temperaments must not be confused with the concept of ``stretched tuning'' in the context of piano tuning which is a completely different story. In pianos, the stiffness of strings causes a deformation of the harmonic spectrum, known as inharmonicity, which is compensated in the tuning process by imposing a stretched tuning curve. Contrarily, in the present work we discuss an \textit{intrinsic} stretch of the temperament itself on the basis of perfectly harmonic sounds (iH=0). 

The paper is organized as follows. In the next section various definitions and notations are introduced. The third section discusses the existing stretched ETs in more detail. After reviewing the concept of entropy in Sect. 4, the model for generating harmonic power spectra is introduced  in Sect. 5. Finally,  the numerical results are presented in Sect. 6, followed by a simple visual interpretation of the results.

\section{Definitions and notations}

\subsection{Power spectra}
Tones generated by musical instruments are composed of many Fourier components, called \textit{partials}. The partials are enumerated by an index $n=1,2,\ldots$ with associated frequencies $f_n$ in ascending order, where the lowest frequency $f_1$ is referred to as the \textit{fundamental frequency} of the tone. 

The power spectral density of a tone is defined as $P(f) = |\tilde\psi(f)|^2$, where $\tilde\psi(f)$ is the Fourier transform of the sound wave $\psi(t)$. In the power spectrum the partials appear as peaks whose area $P_n$ is the power of the $n^{\rm{th}}$ partial. The total power is then given by
\begin{equation}
P_{\rm tot} \;=\; \int_0^\infty \d f \, P(f) \;=\; \sum_{n=1}^{\infty} P_n\,.
\end{equation}
Typically the power of the $n^{\rm th}$ peak $P_n$ varies with increasing $n$ and eventually goes to zero so that the number of partials contributing to the sound is essentially finite~\cite{Koenig}. 

As in the Heisenberg energy-time uncertainty relation, the width of the peaks is inversely proportional to the temporal duration of the sound. In the limit of an infinitely long-lasting tone one obtains an idealized spectrum with infinitely sharp peaks of the form
\begin{equation}
\label{DeltaSpectrum}
P(f) \;=\; \sum_{n=0}^\infty P_n \, \delta (f-f_n), 
\end{equation}
where $\delta$ denotes the Dirac delta function. 

\subsection{Harmonic spectra and pure intervals}

A tone is called \textit{perfectly harmonic} if the frequencies of the partials are exactly given by integer multiples of the fundamental frequency:
\begin{equation}
\label{LinearLaw}
f_n= n f_1\,.
\end{equation}
This means that perfectly harmonic sound are exactly periodic in the fundamental frequency $f_1$, causing a clean and static sound. 

The linear law (\ref{LinearLaw}) holds only for ideal oscillators such as infinitely flexible strings. Realistic oscillators in musical instruments such as piano strings are of course not perfectly harmonic. In the present work, however, we will neglect this instrument-dependent imperfectness, restricting ourselves to the ideal case of perfectly harmonic spectra of the form (\ref{LinearLaw}).

An interval between two fundamental frequencies $f_1$ and $f_1'$ is called \textit{pure} if some of the corresponding higher partials coincide, meaning that there exist integers $n,m \in\mathbb{N}$ such that $f_n= f'_m$. In the case of ideal harmonic oscillators the fundamental frequencies of a pure interval are therefore related by a rational number:
\begin{equation}
nf_1 = mf'_1 \qquad \Rightarrow \qquad \frac{f'_1}{f_1} = \frac{n}{m}\,.
\end{equation}
Since these coincidences appear periodically in frequency space, the perception of harmonicity is particularly intense if $n$ and $m$ are small. Examples are the pure octave (2:1), the perfect fifth (3:2), and the perfect fourth (4:3).

\subsection{Temperaments}
Let us consider a chromatic scale of $K$ semitones with the fundamental frequencies $f_1^{(k)}$ enumerated by $k=1 \ldots K$. The set of all frequencies $\{f_1^{(k)}\}$ with respect to a reference tone (usually A4 = 440 Hz) is referred to as the \textit{temperament} of the musical scale.

A temperament is called \textit{equal} if the frequency ratios of all intervals are invariant under transposition (translational shifts along the keyboard), i.e.,
\begin{equation}
\frac{f_1^{(k+s)}}{f_1^{(k'+s)}} = \frac{f_1^{(k)}}{f_1^{(k')}} \qquad \forall k,k',s\,.
\end{equation}
Obviously this requires all semitone intervals
\begin{equation}
S_k \;:=\; \frac{f_1^{(k+1)}}{f_1^{(k)}}
\end{equation}
to be constant. 

The standard twelve-tone \textit{equal temperament} (ET), which was originally invented in ancient China~\cite{China} and rediscovered in Europe in the \nth{18}th century, is determined by two additional conditions. Firstly the octave is divided into twelve semitones. Secondly the octave, the most fundamental of all intervals, is postulated to be pure (beatless), as described by the frequency ratio 2:1. These two conditions unambiguously imply that the constant semitone frequency ratio is given by $S=2^{1/12}$. Defining a reference tone with the index $k_{\rm ref}$, for example A4 with the frequency $f_{\rm ref}=f_1^{(k_{\rm ref})}=440$ Hz, the standard ET is therefore defined by
\begin{equation}
f_1^{(k)} \;=\; f_{\rm ref} S^{k-k_{\rm ref}} \;=\; f_{\rm ref} 2^{(k-k_{\rm ref})/12}\,.\qquad\qquad\mbox{(ET)}
\end{equation}

\subsection{Pitches}
Since intervals are defined by frequency \textit{ratios} rather than differences, they give rise to a multiplicative structure in the frequency space. It is therefore meaningful and convenient to consider the logarithm of the frequencies, mapping it to an additive structure. In music theory this is done by defining the so-called \textit{pitch}
\begin{equation}
\chi \;:=\; 1200\,\log_2\frac{f}{f_{\rm ref}}\,\qquad \Leftrightarrow \qquad f=f_{\rm ref} \, 2^{\chi/1200}
\end{equation}
with respect to a given reference frequency $f_{\rm ref}$, where $\log_2 z = \ln z/\ln 2$ denotes the logarithm to base 2. With this definition it is possible to translate frequency ratios into pitch differences. In music theory such pitch differences are usually measured in \textit{cents} ($\cent$), defined as 1/100 of a semitone in the standard ET. Therefore, in terms of pitch variables the ET is simply given as
\begin{equation}
\chi_1^{(k)} = 100 (k-k_{\rm ref})\,.\qquad\qquad\mbox{(ET)}
\end{equation}
With pitches it is possible to specify interval sizes conveniently in cents. For example, the perfect fifth has a size of $1200 \log_2\frac32 \approx 701.955\cent$, which is slightly larger than the fifth in the standard ET with a size of exactly $700\cent$.

When going from frequencies to pitches, the power spectral density $P(f)$ has to be recast as a spectral density $P(\chi)$ in terms of the pitch variable by means of the transformation
\begin{equation}
P(\chi) \;=\; \frac{P(f)}{\d\chi/ \d f} \;=\; \frac{1200}{\ln 2}\,\frac{P(f)}{f}\,.
\end{equation}
If we apply this transformation to the idealized peaked spectrum in Eq.~(\ref{DeltaSpectrum}) we find that formally the same expression holds for the pitch variables as well, namely
\begin{equation}
\label{DeltaSpectrum2}
P(\chi) \;=\; \sum_{n=0}^\infty P_n \, \delta (\chi-\chi_n), 
\end{equation}
%

\section{Stretched equal temperaments}

In what follows we consider equal chromatic temperaments of the form
\begin{equation}
f_1^{(k)} \;\propto\; S^k\,,
\end{equation}
where the semitone ratio $S$ is constant but slightly different from $2^{1/12}$, meaning that the semitone interval size deviates marginally from $100\cent$. This deviation, measured in units of cents, will be denoted by $\epsilon$, i.e., we consider equal temperaments with the semitone pitch difference
\begin{equation}
\chi_S \;=\; 100\cent + \epsilon\,.
\end{equation}
The corresponding frequencies and pitches of these modified temperaments are given by
\begin{equation}
\label{frequencies}
f_1^{(k)} \;=\; f_{\rm ref}\, 2^{(k-k_{\rm ref})(1+\epsilon/100)/12}\,,
\end{equation}
\begin{equation}
\label{pitches}
\chi_1^{(k)} \;=\; \chi_{\rm ref}  +  (100+\epsilon)(k-k_{\rm ref})\,.
\end{equation}
Here $\epsilon$ is a free parameter which quantifies the stretch or quench of the temperament relative to the standard ET. As we will see, the stretch parameter $\epsilon$ is of the order of a few hundredths cents. Therefore, as a convenient notation for what follows, we shall denote by ET$_x$ a stretched equal temperament with a semitone stretch of $\epsilon=\frac{x}{1000}\cent$. For example, ET$_0$ is just the standard equal temperament while ET$_{50}$ would denote a stretched equal temperament with $\epsilon = 0.05\cent$ or 50 millicents. Since stretch differences of less than a millicent would not be audible, the resolution in terms of millicents is more than sufficient.

The class of stretched 12-tone ETs comprises several special cases, as will be discussed in the following.

\subsection{Special cases}

The \textbf{Stopper{\texttrademark} equal temperament}~\cite{Stopper} replaces the constraint of pure octaves by pure duodecimes, meaning that $S^{19}=3$. This corresponds to taking
\begin{equation}
\epsilon_{\rm Stopper} \;=\; \frac{1200}{19}\log_2{3}-100 \;\approx \; 0.103\,.
\end{equation}
The stretch of the \textbf{Cordier equal temperament}~\cite{Cordier}, which postulates perfect fifths, is even higher, namely
\begin{equation}
\epsilon_{\rm Cordier} \;=\; \frac{1200}{7}\log_2{\frac32}-100 \;\approx \; 0.279\,.
\end{equation}
Finally, in the \textbf{c.ha.s{\texttrademark} equal temperament} the semitone ratio is defined by the implicit equation~\cite{Capurso,Chiriano}
\begin{equation}
S = (3-\Delta)^{1/19} = (4+s\Delta)^{1/24}
\end{equation}
with two construction-specific parameters $s$ and $\Delta$, where the special case of c.ha.s corresponds to setting $s=1$. Solving this equation one obtains $\Delta \approx 0.0213$, corresponding to the stretch parameter
\begin{equation}
\epsilon_{\rm c.ha.s} = \frac{1200}{19} \log_2(3-\Delta) \;\approx\; 0.038\,.
\end{equation}
All these temperaments, which are summarized in Table \ref{TableET}, are strictly equal in the sense that the semitone frequency ratio is constant. They belong to a whole continuum of possible stretched twelve-tone equal temperaments labeled by the parameter $\epsilon$. 
\begin{table}
\begin{center}
\begin{tabular}{|l|l||l|l|} \hline
Temperament & shortcut & $\epsilon$ & $S$ \\ \hline
Standard ET & ET$_0$ & 0        & 1.0594631 \\
c.ha.s & ET$_{38}$ & 0.038        & 1.0594865 \\
Stopper & ET$_{103}$ & 0.103    & 1.0595251 \\
Cordier & ET$_{279}$ & 0.279    & 1.0596340 \\ \hline
\end{tabular}
\end{center}\caption{Chromatic equal temperaments compared in the present work.
\label{TableET}}
\end{table}

\section{Entropy as a harmonicity measure}

\subsection{Entropy detecting overlapping peaks}

\begin{figure}
\centering\includegraphics[width=\textwidth]{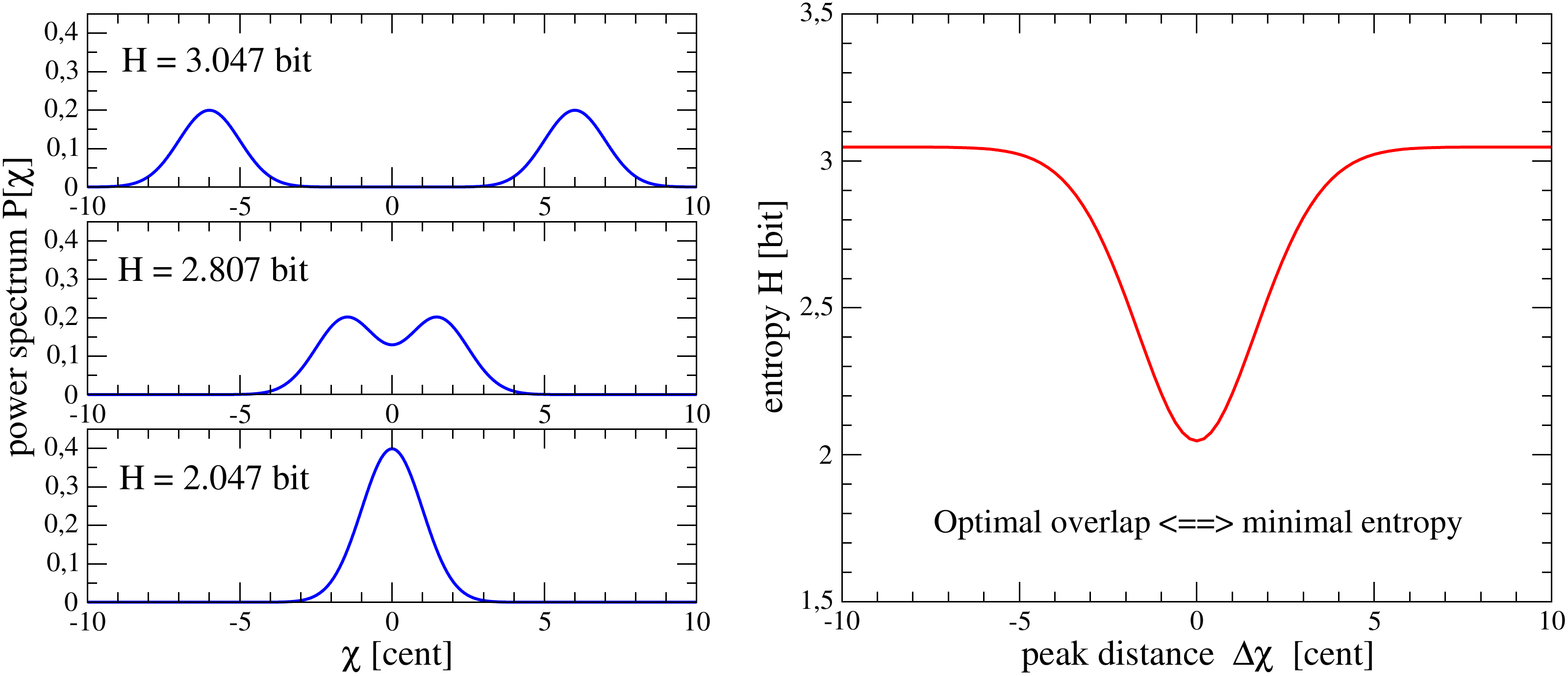}
\caption{\scriptsize Example illustrating how entropy responds to overlapping spectral peaks. The left figure shows a superposition of two peaks with a standard deviation of one cent. If the two peaks do not overlap the corresponding entropy is independent of their distance. However, as soon as the peaks begin to overlap the entropy decreases (see right panel) and becomes minimal for perfect coincidence. This property allows entropy to be used as a measure for the proximity of spectral lines in the power spectrum of sound waves.}
\label{entropydemo}
\end{figure}

In statistical physics and information theory, the differential \textit{entropy} of a normalized probability density $p(\chi)$ is defined as
\begin{equation}
H = - \int \d \chi \,  p(\chi) \log_2 p(\chi)\,.
\end{equation}
The entropy quantifies the information in units of \textit{bits} that is required to specify a randomly chosen value of $x$ according to the distribution $p(x)$ in a given resolution. Thus the enropy of a strongly peaked distribution is low while a broad distribution has a high entropy. This can be demonstrated nicely in the example of a Gaussian distribution
\begin{equation}
\label{NormalDistribution}
p_\sigma(\chi) = \frac{1}{\sigma\sqrt{2\pi}} \, e^{-\frac{\chi^2}{2\sigma^2}}
\end{equation}
for which the entropy
\begin{equation}
\label{EntropyNormalDistribution}
H_\sigma=\frac12 \log_2(2\pi e) + \log_2 \sigma
\end{equation}
increases logarithmically with the standard deviation $\sigma$. But the entropy is not only able to quantify the peakedness of a distribution and its associated information content, even more important in the present context is its ability to detect overlapping peaks. As an example let us consider the sum of two normalized Gaussian peaks of width $\sigma$ which are separated by the distance $\Delta \chi$:
\begin{eqnarray}
p(\chi) &=& \frac12\Bigl[p_\sigma(\chi+\Delta\chi/2) \,+\, p_\sigma(\chi-\Delta\chi/2)\Bigr] \\  &=& \frac{1}{2\sigma\sqrt{2\pi}}  \Bigl( e^{-\frac{(\chi+\Delta\chi/2)^2}{2\sigma^2}} \,+\, e^{-\frac{(\chi-\Delta\chi/2)^2}{2\sigma^2}} \Bigr)\,. \nonumber
\end{eqnarray}
If the two peaks are sufficiently separated like in the upper left panel of Fig.~\ref{entropydemo}, it is clear that the entropy will be independent of their distance $\Delta \chi$. But as soon as the peaks begin to overlap the entropy decreases and becomes minimal for perfect coincidence $\Delta \chi=0$, as shown in the right panel of the figure. It is this property that allows entropy to detect overlapping partials as a signature of consonance and harmony.

\subsection{Entropy applied to acoustic power spectra}

As already outlined in the introduction, an interval is perceived as harmonic if the corresponding partials of the constituting tones coincide partially. Therefore, interpreting the normalized power spectrum as a probability density, it is plausible that the entropy will attain a local minimum for an optimal compromise of overlapping partials. More specifically, in order to compute the entropy we first have to determine the total power
\begin{equation}
\label{TotalPower}
P_{\rm tot} \;=\; \int_0^\infty \d\chi P(\chi) \,.
\end{equation}
Defining the normalized power spectrum $p(\chi) := \frac{P(\chi)}{P_{\rm tot}}$, the entropy associated with the sound wave is then given by
\begin{equation}
H[p] \;=\; -\int_0^\infty \d \chi \, p(\chi) \log_2 p(\chi)\,.
\end{equation}
As one can see, in this approach the influence of the partials is automatically weighted by their power. Therefore, the entropy does not only detect overlapping partials but even assigns to them a different weight according to their intensity.

In the following section we are going to define a certain model for the composition of the power spectrum based on a given temperament. This model is then used to compute the entropy as a function of the parameter $\epsilon$. The goal is to find the value of~$\epsilon$ for which the entropy becomes minimal, indicating the best compromise of overlapping partials and therewith the most harmonic temperament.

\section{Definition of the model}

In the analysis to be presented below we consider an artificial superposition of perfectly harmonic tones. In order to apply the concept of entropy, we need to make two assumptions:
\begin{itemize}
\item We have to make a reasonable assumption about the power of the partials $P_n^{(k)}$. These coefficients determine the texture of a harmonic sound. In realistic situations the power can be distributed over dozens of partials and the coefficients may even vary with time~\cite{Koenig}. However, in order to keep the analysis as simple as possible we shall assume an exponential decrease of the form
\begin{equation}
\label{exponential}
 P_n^{(k)} \;:=\; P_1^{(k)}\,e^{- (n-1)/\lambda} \;\propto \; e^{-n/\lambda}\,,
\end{equation}
where $\lambda$ is a free parameter. This parameter controls the brilliance (overtone richness) of the sound waves. Roughly speaking $\lambda$ can be interpreted as the index of the partial where the center of the cumulative power distribution is located.
\item 
The Dirac $\delta$-peaks in Eq.~(\ref{DeltaSpectrum}) are infinitely focused, meaning that they never overlap. Thus, in order to apply the entropy criterion, we have to give them a finite width. To this end we convolve the spectrum with a normal distribution of constant width~$\sigma$, obtaining the spectrum
\begin{equation}
\label{ConvolvedHarmonicSpectrum}
P^{(k)}(\chi) \;=\; \frac{1}{\sigma\sqrt{2\pi} } \sum_{n=0}^\infty P^{(k)}_n e^{-\frac{(\chi-\chi_n^{(k)})^2}{2\sigma^2}}\,.
\end{equation}
The standard deviation $\sigma$ in cents enters as another free parameter which expresses to what extent our sense of hearing is willing to tolerate deviations from pure intervals. Since the standard ET comes with deviations from pure tuning ranging from 2 to 16 cents, it is reasonable to choose $\sigma$ as a constant in the same range. 
\end{itemize}

\begin{figure}
\centering\includegraphics[width=135mm]{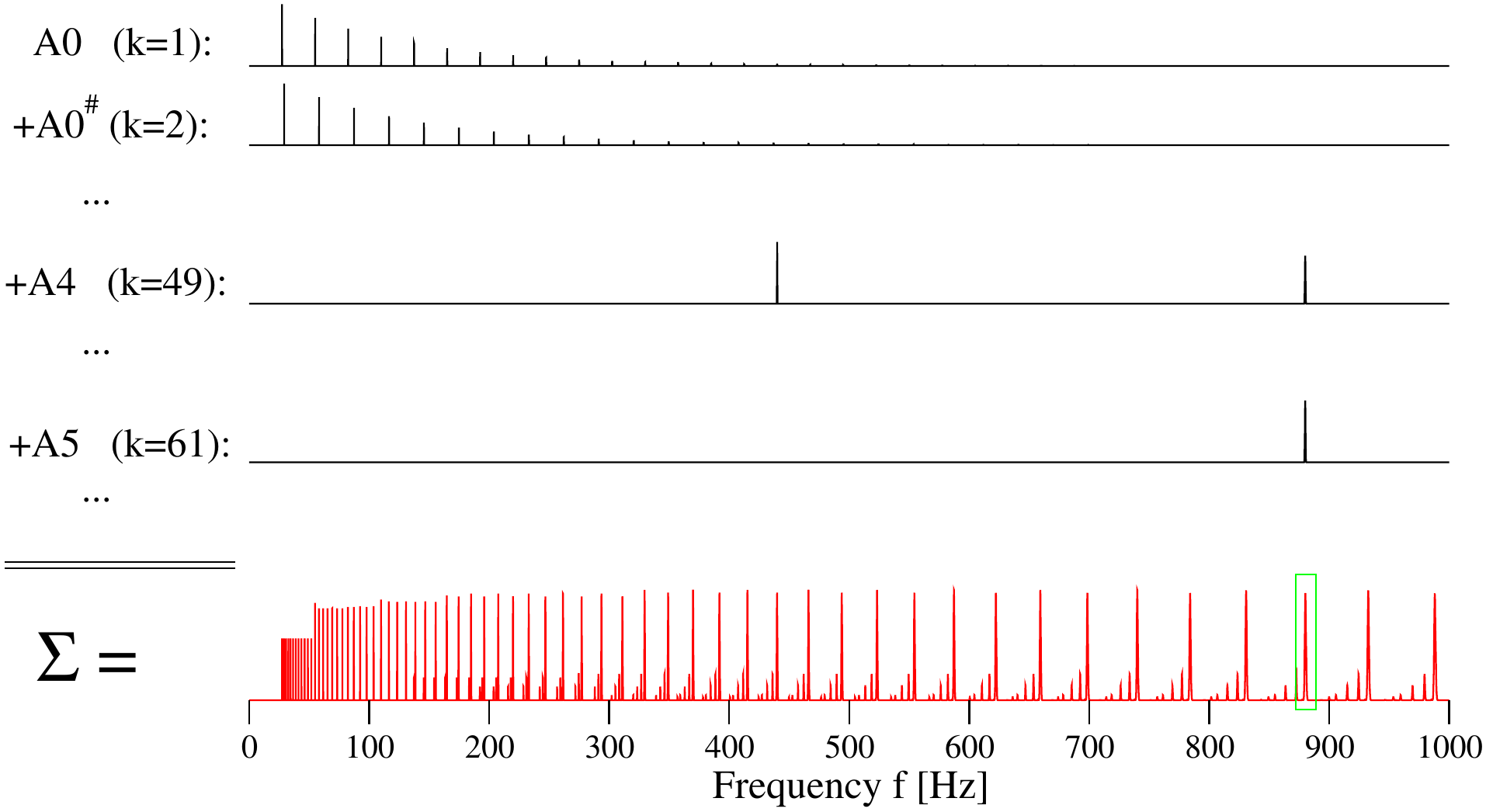}
\caption{\scriptsize Artificially generated power spectra used in the present paper. The black lines represent perfectly harmonic power spectra of $K=88$ tones (four of which are shown) with partials whose power decreases exponentially as $e^{-n/4}$. The sum of all $K$ tones is shown at the bottom in red color. This sum spectrum, which may be thought of as hearing all tones simultaneously, is used to compute the entropy.}
\label{model}
\end{figure}

\noindent
Finally we have to decide which of the tones we would like to compare. In order to avoid any preference the simplest choice is to compare all tones simultaneously, a concept which also turned out to be useful in the context of entropy-based tuning~\cite{EPTPaper}. This means that we simply add up the power spectra of all $K$ tones, as sketched in Fig.~\ref{model}. The resulting power spectrum reads
\begin{equation}
P(\chi) =  \frac{1}{K \sigma \sqrt{2\pi} } \sum_{k=1}^K\sum_{n=0}^\infty P^{(k)}_n e^{-\frac{(\chi-\chi_n^{(k)})^2}{2\sigma^2}} \,.
\end{equation}
Inserting Eq.~(\ref{exponential}) as well as the pitches of the harmonics in the $\epsilon$-stretched temperament 
\begin{equation}
\chi_n^{(k)} \;=\; \chi_{\rm ref}  +  (100+\epsilon)(k-k_{\rm ref})+1200 \log_2 n\,
\end{equation}
and dividing by the total power (\ref{TotalPower}) we arrive at the normalized power spectrum
\begin{equation}
\label{NormalizedProbabilty}
\fl 
p_{\epsilon,K,\lambda,\sigma}(\chi) \;=\; \frac{1}{P_{tot}}\sum_{k=1}^K\sum_{n=0}^\infty \exp\Bigl( -\frac{n}{\lambda} - \frac{(\chi- \chi_{\rm ref} - (100+\epsilon) (k-k_{\rm ref}) - 1200 \log_2 n)^2}{2\sigma^2} \Bigr)
\end{equation}
of which we will compute the entropy
\begin{equation}
\label{EntropyToBeEvaluated}
H_{\epsilon,K,\lambda,\sigma} \;=\; -\int_0^\infty \d \chi \,\,
p_{\epsilon,K,\lambda,\sigma}(\chi) \log_2 
p_{\epsilon,K,\lambda,\sigma}(\chi) \,.
\end{equation}
This entropy depends on four parameters, namely, the stretch parameter $\epsilon$, the number of tones $K$, the overtone richness parameter $\lambda$, and the width of the spectral peaks $\sigma$. Note that $k_{\rm ref}$ and $\chi_{\rm ref}$ cancel out in the normalization so that the entropy does not depend on the overall pitch and the chosen reference key.

\section{Numerical results}

\begin{figure}
\centering\includegraphics[width=120mm]{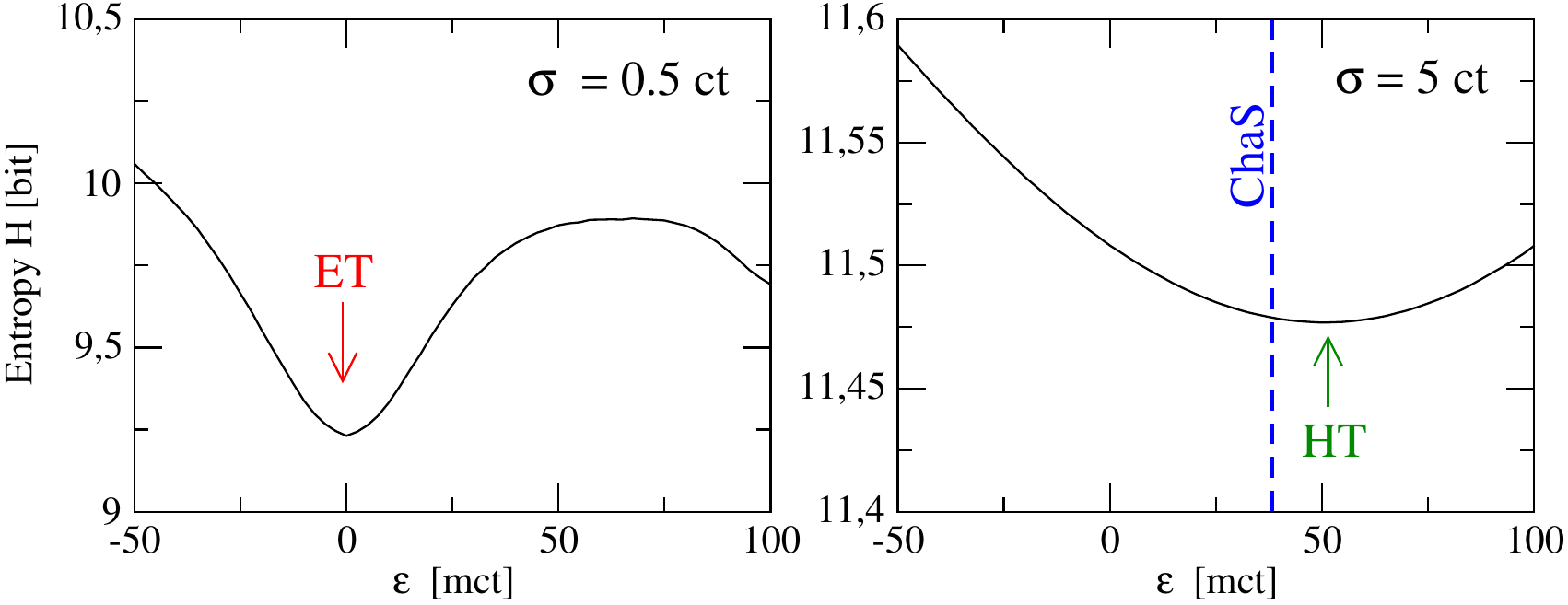}
\caption{\scriptsize Entropy of the generalized temperament as a function of the parameter $\epsilon$ in units of millicents $(\frac{1}{1000}\cent)$ for $K=88$ and $\lambda=10$. For very small values of $\sigma$ (left panel) the entropy is minimal in the standard equal temperament (ET), while for larger values of $\sigma$ (right panel) the minimum (harmonic temperament, HT) is clearly found at a value $\epsilon>0$. The value is slightly larger than the one predicted in Ref.~\cite{Capurso} (dashed line) but it is of the same order of magnitude.
}
\label{example}
\end{figure}

The integral (\ref{NormalizedProbabilty})-(\ref{EntropyToBeEvaluated}) is not exactly solvable and thus it can only be calculated numerically. Moreover, the integrand is highly oscillatory so that ordinary integrators of algebraic computer systems are prone to fail. Therefore, we compute the integral by an ordinary Simpson integration in C++, discretizing the pitch $\chi$ in bins of 1/1000 cents. 

For a first survey we consider a particular example which is shown in Fig.~\ref{example}. In this example we chose $K=88$ tones and the decay parameter $\lambda=10$. The left panel shows the entropy as a function of $\epsilon$ for $\sigma=0.5\cent$ which is much smaller than the tolerance of human hearing. In this case the entropy is minimal for $\epsilon=0$, which is just the standard ET. This result is plausible since for $\epsilon=0$ the octaves lock in and the entropy measure is so rigid that it does not like to depart from this point. However, as we increase the tolerance, e.g. by taking the realistic value $\sigma=5\cent$ as in the right panel, we find as expected that the entropy changes less, but most strikingly there is a new minimum appearing at a non-vanishing value of $\epsilon$. This means that the corresponding stretched temperament should be more harmonic than the standard ET. The minimum is located at $\epsilon \approx 0.05\cent$, somewhat larger than the value $\epsilon=0.038 \cent$ predicted by the c.ha.s but significantly lower than the values suggested by Stopper and Cordier.

\begin{figure}
\centering\includegraphics[width=100mm]{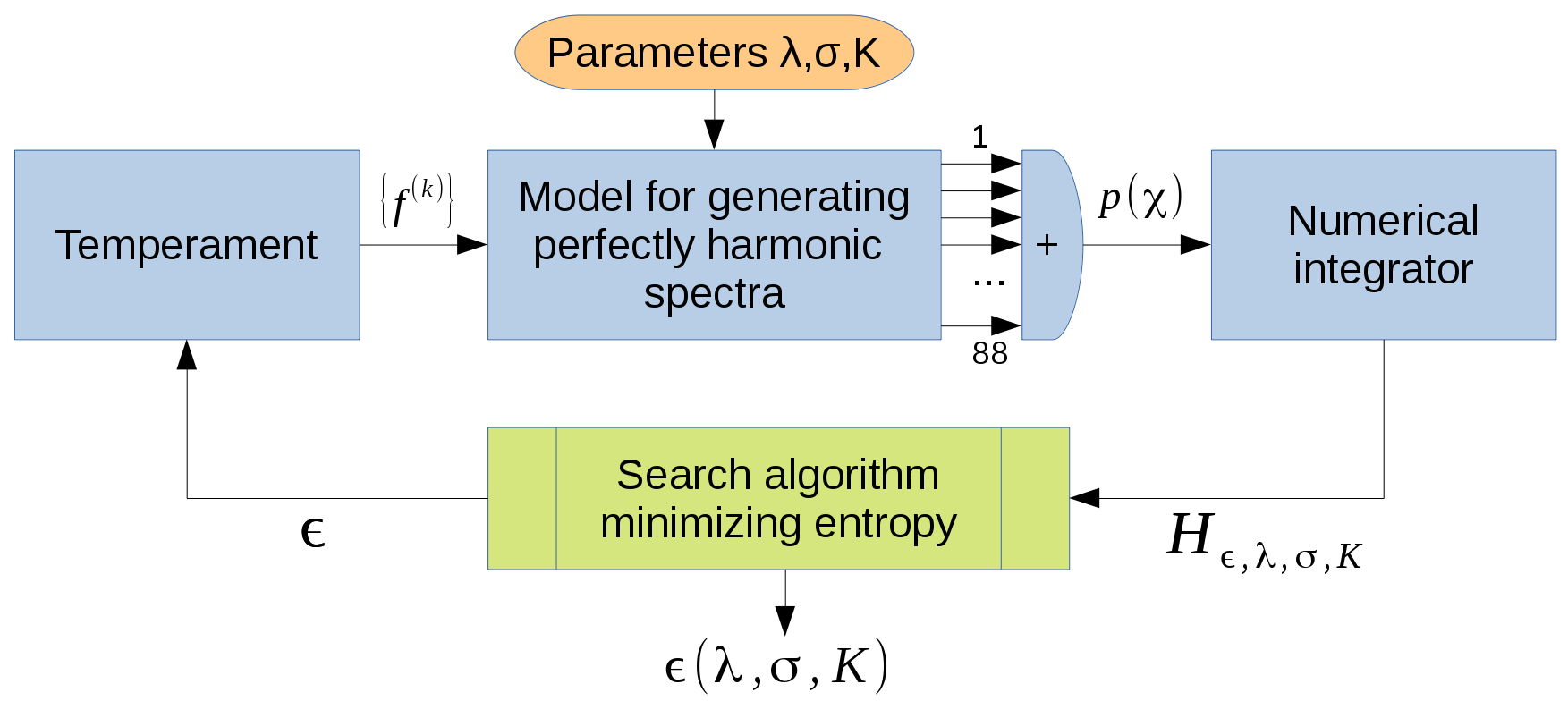}
\vspace*{-3mm}
\caption{\scriptsize Schematic illustration of the entropy minimization method.
}
\label{method}
\end{figure}
\begin{figure}
\centering
\includegraphics[width=85mm]{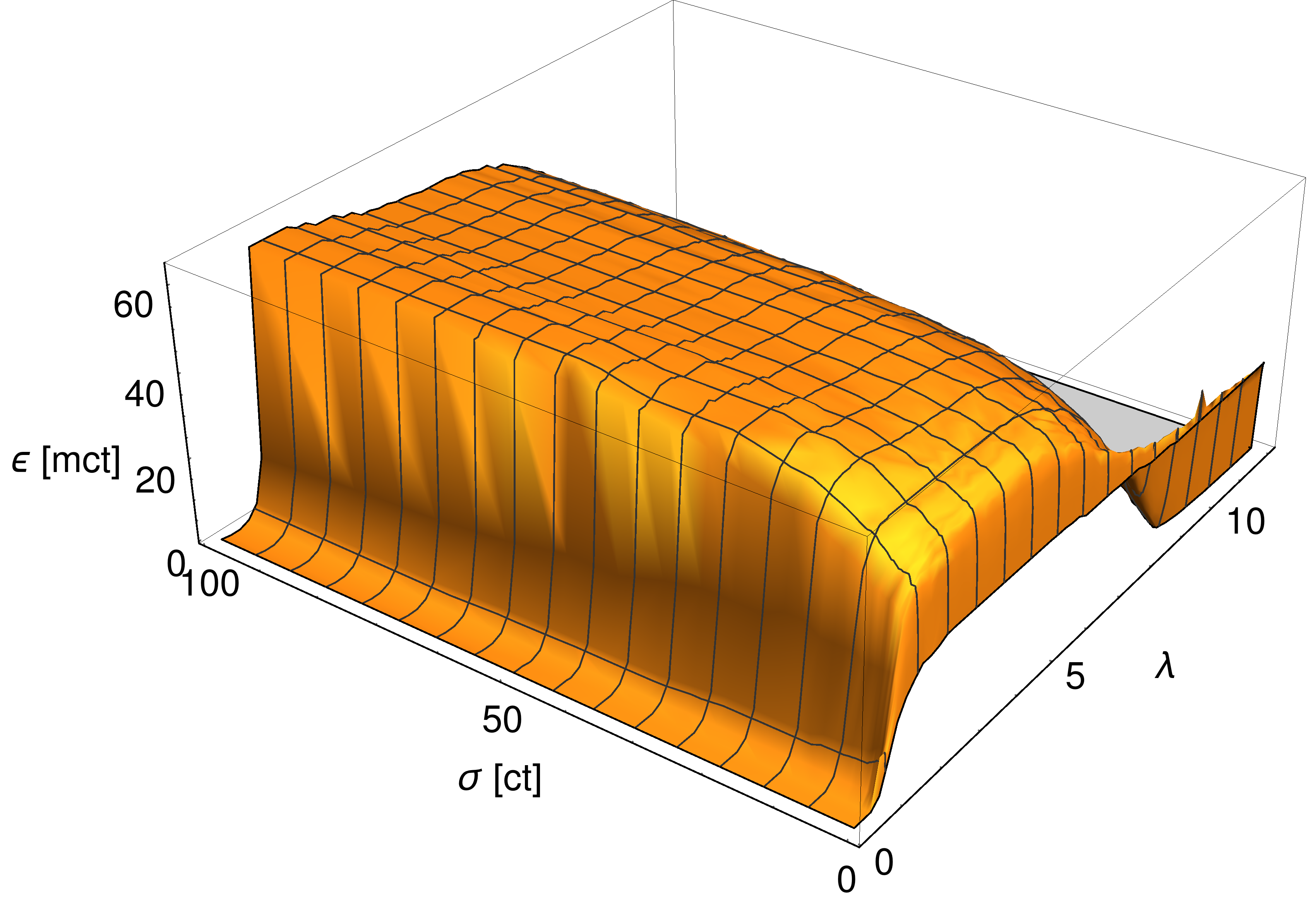}
\includegraphics[width=69mm]{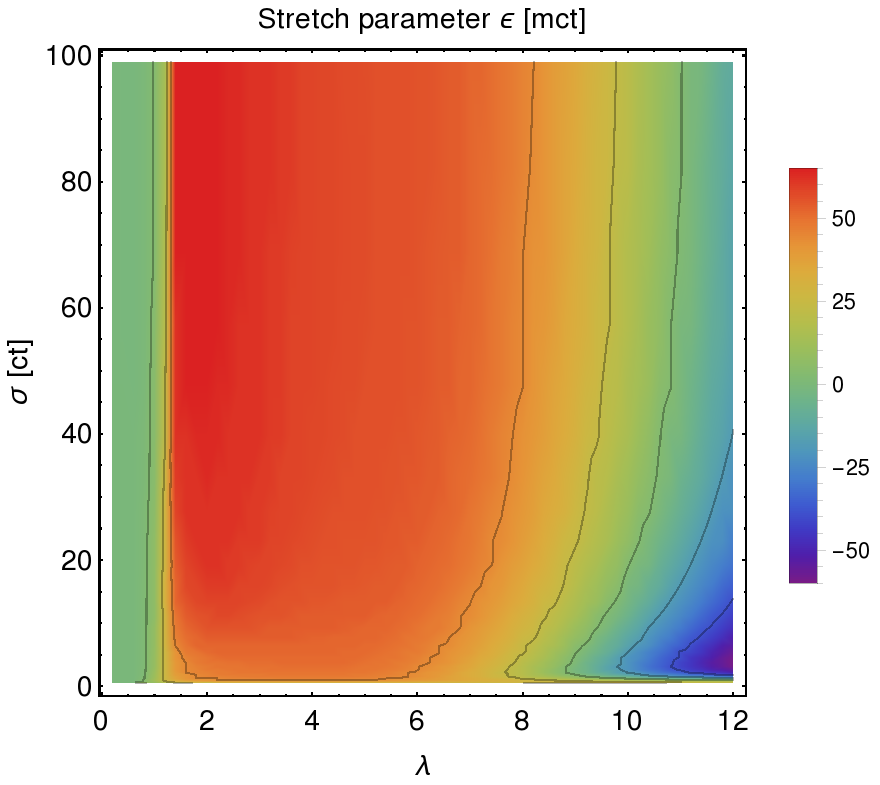}
\caption{\scriptsize Systematic scan over the $(\lambda,\sigma)$ parameter space for a scale with $K=88$ tones. Both figures show the optimal value of  parameter~$\epsilon$, where the entropy becomes minimal, in units of millicents (mct). As can be seen, an extended plateau emerges where $\epsilon$ is rather insensitive to the parameters, indicating a high degree of robustness.
}
\label{scan}
\end{figure}

As we have made various additional assumptions in our model expressed in terms of the parameters $\sigma$ and $\lambda$, we have to analyze the robustness of this result. To this end we carried out a systematic scan over the whole range of $\lambda$ and $\sigma$. For given values of these parameters we search for the value of $\epsilon$ where the entropy $H$ becomes minimal, as sketched in Fig.~\ref{method}. The result of this scan (see Fig.~\ref{scan}) can be interpreted as follows:
\begin{itemize}
\item
\textbf{Parameter $\mathbf\lambda$ controlling the overtone richness:} For $\lambda=0$ (plain sine waves without higher partials) temperaments have no meaning. For small $\lambda<1$ only the fundamental and the second partial contribute. For this reason octaves lock in so that the standard ET is favored. However, as $\lambda$ increases so that more and more partials contribute, the minimum suddenly jumps to a high value of $\epsilon \approx 0.06\cent$. From here on the entropy method clearly prefers a stretched temperament. This value is pretty stable until for very large $\lambda>6$ one observes a gradual decrease, where so many partials begin to contribute that the method becomes unstable.
\item 
\textbf{Parameter $\mathbf\sigma$ controlling the peak width:} For $\sigma=0$ the method does not tolerate any beats, hence the standard temperament with pure octaves provides a local minimum. As shown before, the situation changes suddenly when $\sigma$ exceeds a threshold of about $2 \cent$. Moreover, the figure shows that the optimal value of $\epsilon$ is quite stable as we increase~$\sigma$ even further. In other words, increasing our tolerance for pitch deviations does not lead to significantly different solutions for the optimal temperament.
\end{itemize}
The extended plateau shows that the two ad-hoc parameters, the spectral brilliance expressed by $\lambda$ and the human pitch mismatch tolerance expressed by $\sigma$, have only little influence on $\epsilon$, suggesting that the result is stable and robust. On the plateau the optimal value of $\epsilon$ ranges roughly between $0.040\cent$ and $0.065\cent$. 

Finally we studied the influence of the parameter $K$, the total number of semitones in the scale (not shown here). Taking $K=49$ we basically find the same structure although the level of the plateau is somewhat lower, ranging approximately from $0.035\cent$ to $0.055\cent$, now including c.ha.s as a special case.

\section{Visual interpretation}

\begin{figure}
\centering
\includegraphics[width=130mm]{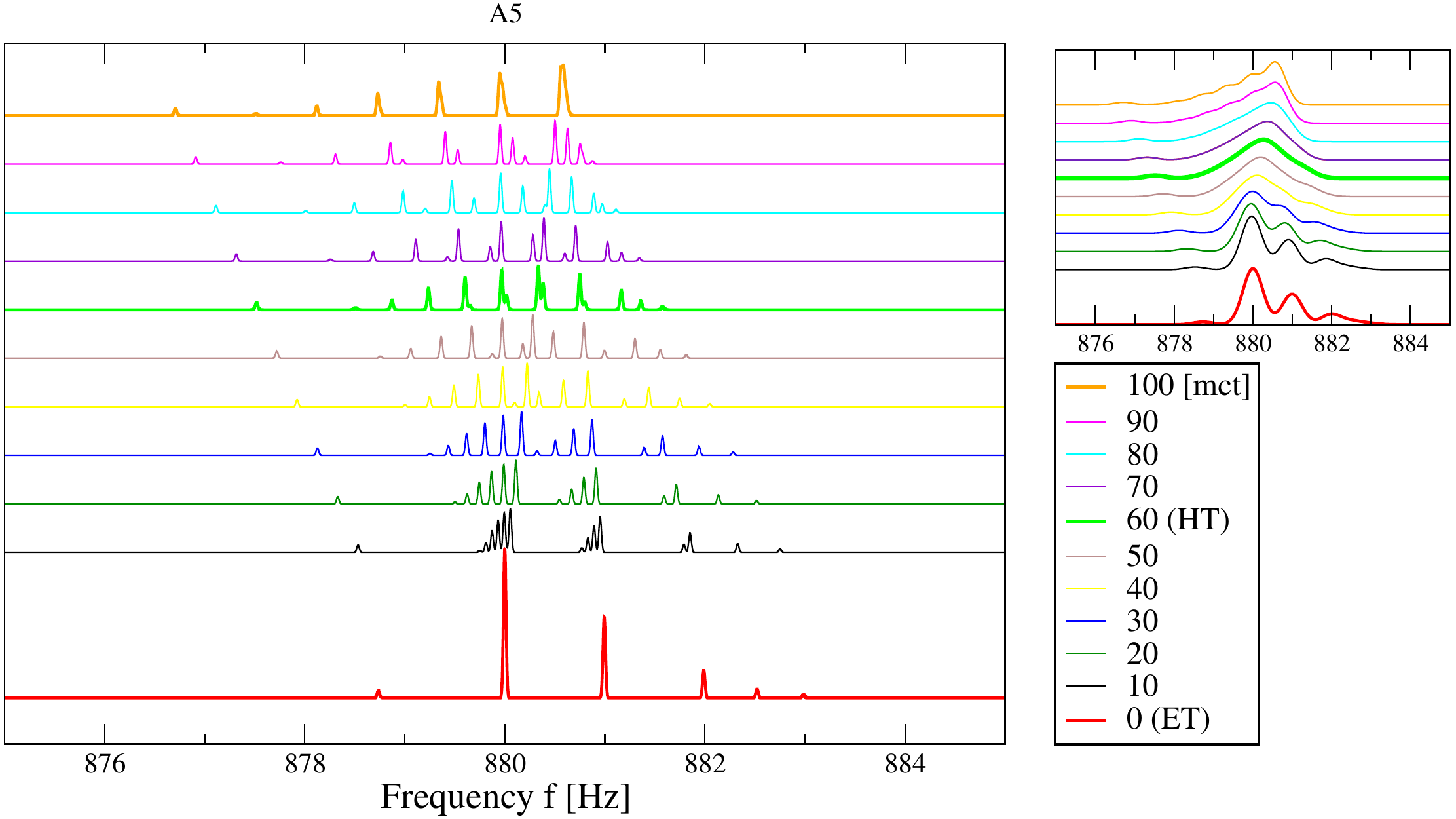}
\caption{\scriptsize Understanding the effect of stretching visually. Left panel: Variation of the spectral sum (\ref{EntropyToBeEvaluated}) near the fundamental frequency of A5 for various values of the stretch parameter $\epsilon$ (see legend). Upper right panel: Same data shown for spectral peaks with the standard deviation $\sigma=2$.
}
\label{visual}
\end{figure}

The results obtained in the previous section can be understood quite easily by visual inspection of the spectral sum. To this end we zoom the power spectrum in a small corridor around 880 Hz, corresponding to the tone A5 (highlighted as a small green rectangle in Fig.~\ref{model}). In the left panel of Fig.~\ref{visual} we demonstrate how the spectral lines around A5 depend on the stretch parameter $\epsilon$. Our observations can be summarized as follows:
\begin{itemize}
 \item In case of the equal temperament (red line at the bottom) the sound of A5 consists of five peaks. The peak at 880 Hz is the strongest one because it is made up of six contributions coming from A0,A1,...,A5 by means of pure octaves. Moreover, we observe four adjacent peaks on the right side which are generated by different interval combinations on the chromatic scale. Remarkably, these peaks are arranged in a totally asymmetric way, i.e., they are found \textit{exclusively on the right side} of the fundamental peak.
 \item Choosing a small value $\epsilon>0$ the octave symmetry is broken, lifting the degeneracy of the peaks. Increasing $\epsilon$ we finally arrive at a point of maximal symmetry at $\epsilon \approx 0.05\cent$. At this point one has small peaks which are arranged symmetrically around 880 Hz.
 \item Increasing the stretch parameter even further another interval is becoming pure, leading again to asymmetrically distributed degenerated peaks. This is just the Stopper temperament, favoring pure duodecimes.
\end{itemize}
The figure gives also an intuitive understanding why the stretched temperament for $\epsilon \approx 0.05\cent$ is more harmonic and why the entropy is able to detect the harmonic optimum. This is demonstrated in the upper right panel, where the same data is shown for a broader peaks with the width $\sigma=2$. As can be seen by bare eye, the green curve in the middle is the most symmetric and most focused one, making it plausible why entropy becomes minimal at this point.

\section{Hearing the difference}

Within a single octave the tiny stretch differences discussed in the present paper are so small that they are probably not audible. However, the situation is different for composite sounds extending over several octaves. Here the dependence on $\epsilon$ can be heard in the balance of beat frequencies of the partials.

As a simple example let us consider a particular chord based on C3, combined with two fifths C4-G4-C5-G5. In the standard ET the partials at C4 and C5 match perfectly without any beats while G4 and G5 differ slightly from the corresponding partials in the harmonic series of C3, exhibit 0.44 and 0.89 beats per second. In the stretched case these beat frequencies are reduced on the expense of the octave which starts to beat by itself. In the optimal range of c.ha.s and ET$_{50}$ all these beats are balanced and well below 0.5 Hz. However, increasing $\epsilon$ even further the situation becomes again worse, as summarized in Table~\ref{TableBeats}.

\begin{table}
\begin{center}
\small
\begin{tabular}{|l||l|l|l|l|l|} \hline
Temperament & Standard & c.ha.s \ \ \ & this work & Stopper & Cordier\\
Shortcut & ET$_{0}$ & ET$_{38}$ & ET$_{50}$ & ET$_{103}$ & ET$_{279}$\\ \hline
beats at C4 & 0 	& 0.07 & 0.09 & 0.18 & 0.5\\
beats at G4 & 0.44 	& 0.28 & 0.23 & 0 & 0.75\\
beats at C5 & 0  	& 0.28  & 0.36 & 0.74 & 2.02\\
beats at G5 & 0.89 	& 0.35 & 0.18 & 0.56 & 3.03 \\ \hline
\end{tabular}
\includegraphics[width=115mm]{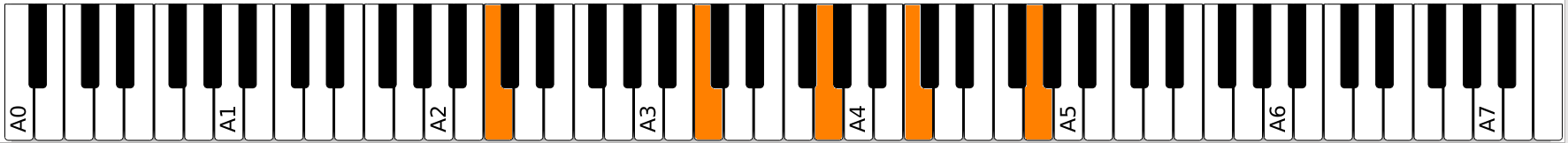}\\
\caption{\scriptsize Frequency of beats in Hz for the chord C3-C4-G4-C5-G5}
\end{center}
\label{TableBeats}
\end{table}

\section{Conclusions}

Using entropy as a measure of harmonicity we have provided numerical evidence that the standard equal temperament with semitone frequency ratio $2^{1/12}$ does \textit{not} represent the harmonic optimum, the optimum is rather obtained for a slightly larger frequency ratio. In such a stretched equal temperament the octave is no longer beatless but by tolerating slightly tempered octaves the beats of other intervals can be reduced, leading all in all to a more harmonic and balanced temperament. The proposed harmonized temperament (HT) is defined by
\begin{equation}
\label{HT}
f^{(k)} \;=\; f_{\rm ref} \, 2^{\frac{(1+\epsilon/100)}{12}(k-k_{\rm ref})}
\end{equation}
with a free parameter $\epsilon$ (see Eqs. (\ref{frequencies})-(\ref{pitches})). Depending on the number of tones in the scale and the richness of partials the method of entropy minimization predicts values in the range 
\begin{equation}
\label{Range}
0.035 \, \cent \;\lesssim\; \epsilon \;\lesssim\; 0.065 \, \cent.
\end{equation}
According to this result, the temperaments proposed by Stopper and Cordier can be ruled out as long as we restrict ourselves to perfectly harmonic sounds. The c.ha.s\texttrademark \ temperament ($\epsilon=0.038\,\cent$), however, is consistent with the entropy method, indicating that c.ha.s is capturing the right idea, namely, to embrace all intervals as part 
of a whole and to look for the optimal order of beats. However, choosing the most plausible parameters the predicted optimal value for $\epsilon$ tends to be somewhat larger, roughly centered around
\begin{equation}
\epsilon \approx 0.05\,\cent\,,
\end{equation}
and it would be interesting to study where this discrepancy comes from.
\texttt{}
Let us again remind the reader that the predicted stretch $\epsilon$ has \textit{nothing} to do with stretched tuning curves in the context of piano tuning. This would be an additional effect on top of the present one caused by the physical inharmonicity of piano strings. Contrarily, in the present study the semitone stretch $\epsilon$ is a property of the equal temperament itself, assuming perfectly harmonic spectra of partials. Nevertheless we expect our findings to be also relevant for piano tuning if inharmonicity is taken into account as an additional effect.

In conclusion our results call for replacing the equal temperament by a stretched equal temperament of the form~(\ref{HT}) as a new standard for musical instruments with fixed frequencies.

An important open question would be the appropriate choice of the parameter~$\epsilon$ for such a new standard. Would it be sufficient to choose a fixed value as a reasonable compromise for all purposes? Or do we have to keep $\epsilon$ as a free parameter varying in the range~(\ref{Range}), chosen as a matter of taste? Careful empirical studies are needed in order to find out how much this parameter influences the perception of the temperament.

\vspace{3mm}

\noindent
\textbf{\large Acknowledgements:}\\[5mm]
I am deeply grateful to \textbf{Alfredo Capurso} (Messina, Italy), creator of c.ha.s\texttrademark \ and renowned expert for tuning and temperaments, for his helpful comments and for carefully reading the manuscript before submission. I would also like to thank \textbf{Isaac Oleg} (Paris, France), who is internationally known as an innovative piano technician, for interesting discussions and for having me made aware of the problem addressed in this paper. Without both of them the present work would not exist.

\vspace{-3mm}


\appendix
\section{Frequency table}

\begin{center}
\begin{scriptsize}\begin{tabular}{|c|c||c|cc|cc|cc|} \hline
$k$ & tone & ET [Hz] & c.ha.s [Hz] & $\Delta\chi \, [\cent]$ & ET$_{50}$ [Hz] & $\Delta\chi \,[\cent]$  & ET$_{60}$ [Hz] & $\Delta\chi \,[\cent]$ \\ \hline
1	& A 0	& 27.50	& 27.47	& -1.82	& 27.46	& -2.40	& 27.45	& -2.88\\
2	& A$^\sharp$0	& 29.14	& 29.11	& -1.79	& 29.10	& -2.35	& 29.09	& -2.82\\
3	& B 0	& 30.87	& 30.84	& -1.75	& 30.83	& -2.30	& 30.82	& -2.76\\
\hline
4	& C 1	& 32.70	& 32.67	& -1.71	& 32.66	& -2.25	& 32.65	& -2.70\\
5	& C$^\sharp$1	& 34.65	& 34.61	& -1.67	& 34.60	& -2.20	& 34.60	& -2.64\\
6	& D 1	& 36.71	& 36.67	& -1.63	& 36.66	& -2.15	& 36.65	& -2.58\\
7	& D$^\sharp$1	& 38.89	& 38.86	& -1.60	& 38.84	& -2.10	& 38.83	& -2.52\\
8	& E 1	& 41.20	& 41.17	& -1.56	& 41.15	& -2.05	& 41.14	& -2.46\\
9	& F 1	& 43.65	& 43.62	& -1.52	& 43.60	& -2.00	& 43.59	& -2.40\\
10	& F$^\sharp$1	& 46.25	& 46.21	& -1.48	& 46.20	& -1.95	& 46.19	& -2.34\\
11	& G 1	& 49.00	& 48.96	& -1.44	& 48.95	& -1.90	& 48.93	& -2.28\\
12	& G$^\sharp$1	& 51.91	& 51.87	& -1.41	& 51.86	& -1.85	& 51.85	& -2.22\\
13	& A 1	& 55.00	& 54.96	& -1.37	& 54.94	& -1.80	& 54.93	& -2.16\\
14	& A$^\sharp$1	& 58.27	& 58.23	& -1.33	& 58.21	& -1.75	& 58.20	& -2.10\\
15	& B 1	& 61.74	& 61.69	& -1.29	& 61.67	& -1.70	& 61.66	& -2.04\\
\hline
16	& C 2	& 65.41	& 65.36	& -1.25	& 65.34	& -1.65	& 65.33	& -1.98\\
17	& C$^\sharp$2	& 69.30	& 69.25	& -1.22	& 69.23	& -1.60	& 69.22	& -1.92\\
18	& D 2	& 73.42	& 73.37	& -1.18	& 73.35	& -1.55	& 73.34	& -1.86\\
19	& D$^\sharp$2	& 77.78	& 77.73	& -1.14	& 77.71	& -1.50	& 77.70	& -1.80\\
20	& E 2	& 82.41	& 82.35	& -1.10	& 82.34	& -1.45	& 82.32	& -1.74\\
21	& F 2	& 87.31	& 87.25	& -1.06	& 87.24	& -1.40	& 87.22	& -1.68\\
22	& F$^\sharp$2	& 92.50	& 92.44	& -1.03	& 92.43	& -1.35	& 92.41	& -1.62\\
23	& G 2	& 98.00	& 97.94	& -0.99	& 97.93	& -1.30	& 97.91	& -1.56\\
24	& G$^\sharp$2	& 103.83	& 103.77	& -0.95	& 103.75	& -1.25	& 103.74	& -1.50\\
25	& A 2	& 110.00	& 109.94	& -0.91	& 109.92	& -1.20	& 109.91	& -1.44\\
26	& A$^\sharp$2	& 116.54	& 116.48	& -0.87	& 116.46	& -1.15	& 116.45	& -1.38\\
27	& B 2	& 123.47	& 123.41	& -0.84	& 123.39	& -1.10	& 123.38	& -1.32\\
\hline
28	& C 3	& 130.81	& 130.75	& -0.80	& 130.73	& -1.05	& 130.72	& -1.26\\
29	& C$^\sharp$3	& 138.59	& 138.53	& -0.76	& 138.51	& -1.00	& 138.50	& -1.20\\
30	& D 3	& 146.83	& 146.77	& -0.72	& 146.75	& -0.95	& 146.74	& -1.14\\
31	& D$^\sharp$3	& 155.56	& 155.50	& -0.68	& 155.48	& -0.90	& 155.47	& -1.08\\
32	& E 3	& 164.81	& 164.75	& -0.65	& 164.73	& -0.85	& 164.72	& -1.02\\
33	& F 3	& 174.61	& 174.55	& -0.61	& 174.53	& -0.80	& 174.52	& -0.96\\
34	& F$^\sharp$3	& 185.00	& 184.94	& -0.57	& 184.92	& -0.75	& 184.90	& -0.90\\
35	& G 3	& 196.00	& 195.94	& -0.53	& 195.92	& -0.70	& 195.90	& -0.84\\
36	& G$^\sharp$3	& 207.65	& 207.59	& -0.49	& 207.57	& -0.65	& 207.56	& -0.78\\
37	& A 3	& 220.00	& 219.94	& -0.46	& 219.92	& -0.60	& 219.91	& -0.72\\
38	& A$^\sharp$3	& 233.08	& 233.03	& -0.42	& 233.01	& -0.55	& 232.99	& -0.66\\
39	& B 3	& 246.94	& 246.89	& -0.38	& 246.87	& -0.50	& 246.86	& -0.60\\
\hline
40	& C 4	& 261.63	& 261.57	& -0.34	& 261.56	& -0.45	& 261.54	& -0.54\\
41	& C$^\sharp$4	& 277.18	& 277.13	& -0.30	& 277.12	& -0.40	& 277.11	& -0.48\\
42	& D 4	& 293.66	& 293.62	& -0.27	& 293.61	& -0.35	& 293.59	& -0.42\\
43	& D$^\sharp$4	& 311.13	& 311.09	& -0.23	& 311.07	& -0.30	& 311.06	& -0.36\\
44	& E 4	& 329.63	& 329.59	& -0.19	& 329.58	& -0.25	& 329.57	& -0.30\\
45	& F 4	& 349.23	& 349.20	& -0.15	& 349.19	& -0.20	& 349.18	& -0.24\\
46	& F$^\sharp$4	& 369.99	& 369.97	& -0.11	& 369.96	& -0.15	& 369.96	& -0.18\\
47	& G 4	& 392.00	& 391.98	& -0.08	& 391.97	& -0.10	& 391.97	& -0.12\\
48	& G$^\sharp$4	& 415.30	& 415.30	& -0.04	& 415.29	& -0.05	& 415.29	& -0.06\\
49	& A 4	& 440.00	& 440.00	& 0.00	& 440.00	& 0.00	& 440.00	& 0.00\\
50	& A$^\sharp$4	& 466.16	& 466.17	& 0.04	& 466.18	& 0.05	& 466.18	& 0.06\\
51	& B 4	& 493.88	& 493.90	& 0.08	& 493.91	& 0.10	& 493.92	& 0.12\\
\hline
52	& C 5	& 523.25	& 523.29	& 0.11	& 523.30	& 0.15	& 523.31	& 0.18\\
53	& C$^\sharp$5	& 554.37	& 554.41	& 0.15	& 554.43	& 0.20	& 554.44	& 0.24\\
54	& D 5	& 587.33	& 587.39	& 0.19	& 587.41	& 0.25	& 587.43	& 0.30\\
55	& D$^\sharp$5	& 622.25	& 622.34	& 0.23	& 622.36	& 0.30	& 622.38	& 0.36\\
56	& E 5	& 659.26	& 659.36	& 0.27	& 659.39	& 0.35	& 659.42	& 0.42\\
57	& F 5	& 698.46	& 698.58	& 0.30	& 698.62	& 0.40	& 698.65	& 0.48\\
58	& F$^\sharp$5	& 739.99	& 740.14	& 0.34	& 740.18	& 0.45	& 740.22	& 0.54\\
59	& G 5	& 783.99	& 784.16	& 0.38	& 784.22	& 0.50	& 784.26	& 0.60\\
60	& G$^\sharp$5	& 830.61	& 830.81	& 0.42	& 830.87	& 0.55	& 830.93	& 0.66\\
61	& A 5	& 880.00	& 880.23	& 0.46	& 880.31	& 0.60	& 880.37	& 0.72\\
62	& A$^\sharp$5	& 932.33	& 932.59	& 0.49	& 932.68	& 0.65	& 932.75	& 0.78\\
63	& B 5	& 987.77	& 988.07	& 0.53	& 988.17	& 0.70	& 988.25	& 0.84\\
\hline
\end{tabular}

\begin{tabular}{|c|c||c|cc|cc|cc|} \hline
$k$ & tone & ET [Hz] & c.ha.s [Hz] & $\Delta\chi \, [\cent]$ & ET$_{50}$ [Hz] & $\Delta\chi \,[\cent]$  & ET$_{60}$ [Hz] & $\Delta\chi \,[\cent]$ \\ \hline
64	& C 6	& 1046.50	& 1046.85	& 0.57	& 1046.96	& 0.75	& 1047.05	& 0.90\\
65	& C$^\sharp$6	& 1108.73	& 1109.12	& 0.61	& 1109.24	& 0.80	& 1109.35	& 0.96\\
66	& D 6	& 1174.66	& 1175.10	& 0.65	& 1175.24	& 0.85	& 1175.35	& 1.02\\
67	& D$^\sharp$6	& 1244.51	& 1245.00	& 0.68	& 1245.16	& 0.90	& 1245.28	& 1.08\\
68	& E 6	& 1318.51	& 1319.06	& 0.72	& 1319.23	& 0.95	& 1319.38	& 1.14\\
69	& F 6	& 1396.91	& 1397.53	& 0.76	& 1397.72	& 1.00	& 1397.88	& 1.20\\
70	& F$^\sharp$6	& 1479.98	& 1480.66	& 0.80	& 1480.88	& 1.05	& 1481.06	& 1.26\\
71	& G 6	& 1567.98	& 1568.74	& 0.84	& 1568.98	& 1.10	& 1569.18	& 1.32\\
72	& G$^\sharp$6	& 1661.22	& 1662.06	& 0.87	& 1662.32	& 1.15	& 1662.54	& 1.38\\
73	& A 6	& 1760.00	& 1760.93	& 0.91	& 1761.22	& 1.20	& 1761.46	& 1.44\\
74	& A$^\sharp$6	& 1864.66	& 1865.68	& 0.95	& 1866.00	& 1.25	& 1866.27	& 1.50\\
75	& B 6	& 1975.53	& 1976.66	& 0.99	& 1977.02	& 1.30	& 1977.31	& 1.56\\
\hline
76	& C 7	& 2093.00	& 2094.25	& 1.03	& 2094.64	& 1.35	& 2094.96	& 1.62\\
77	& C$^\sharp$7	& 2217.46	& 2218.82	& 1.06	& 2219.25	& 1.40	& 2219.61	& 1.68\\
78	& D 7	& 2349.32	& 2350.81	& 1.10	& 2351.29	& 1.45	& 2351.68	& 1.74\\
79	& D$^\sharp$7	& 2489.02	& 2490.66	& 1.14	& 2491.17	& 1.50	& 2491.61	& 1.80\\
80	& E 7	& 2637.02	& 2638.82	& 1.18	& 2639.38	& 1.55	& 2639.86	& 1.86\\
81	& F 7	& 2793.83	& 2795.79	& 1.22	& 2796.41	& 1.60	& 2796.93	& 1.92\\
82	& F$^\sharp$7	& 2959.96	& 2962.10	& 1.25	& 2962.78	& 1.65	& 2963.34	& 1.98\\
83	& G 7	& 3135.96	& 3138.30	& 1.29	& 3139.04	& 1.70	& 3139.66	& 2.04\\
84	& G$^\sharp$7	& 3322.44	& 3324.99	& 1.33	& 3325.80	& 1.75	& 3326.47	& 2.10\\
85	& A 7	& 3520.00	& 3522.78	& 1.37	& 3523.66	& 1.80	& 3524.39	& 2.16\\
86	& A$^\sharp$7	& 3729.31	& 3732.34	& 1.41	& 3733.30	& 1.85	& 3734.10	& 2.22\\
87	& B 7	& 3951.07	& 3954.36	& 1.44	& 3955.41	& 1.90	& 3956.27	& 2.28\\
\hline
88	& C 8	& 4186.01	& 4189.59	& 1.48	& 4190.73	& 1.95	& 4191.67	& 2.34\\
\hline
\end{tabular}

Table 2: Frequency table for various temperaments together with the pitch deviations relative to the standard ET.
\end{scriptsize}
\end{center}

\end{document}